\documentclass[useAMS,usenatbib]{mnras}

\usepackage{color}
\usepackage[english]{babel}
\usepackage{graphicx}
\usepackage{amsmath}
\usepackage{amssymb}
\usepackage{epsf}
\usepackage{bm}
\usepackage[utf8]{inputenc}
\usepackage[T1]{fontenc}
\begin{document}

\def\pFn{p_{\rm Fn}}
\def\pFp{p_{\rm Fp}}
\def\pFe{p_{\rm Fe}}

\newcommand{\om}{\mbox{$\omega$}}              
\newcommand{\Om}{\mbox{$\Omega$}}              
\newcommand{\Th}{\mbox{$\Theta$}}              
\newcommand{\ph}{\mbox{$\varphi$}}             
\newcommand{\del}{\mbox{$\delta$}}             
\newcommand{\Del}{\mbox{$\Delta$}}             
\newcommand{\lam}{\mbox{$\lambda$}}            
\newcommand{\Lam}{\mbox{$\Lambda$}}            
\newcommand{\ep}{\mbox{$\varepsilon$}}         
\newcommand{\ka}{\mbox{$\kappa$}}              
\newcommand{\dd}{\mbox{d}}                     
\newcommand{\nb}{\mbox{$n_{\rm b}$}}                     

\newcommand{\kTF}{\mbox{$k_{\rm TF}$}}         
\newcommand{\kB}{\mbox{$k_{\rm B}$}}           
\newcommand{\tn}{\mbox{$T_{{\rm c}n}$}}        
\newcommand{\tp}{\mbox{$T_{{\rm c}p}$}}        
\newcommand{\te}{\mbox{$T_{eff}$}}             
\newcommand{\ex}{\mbox{\rm e}}                 
\newcommand{\rate}{\mbox{${\rm erg~cm^{-3}~s^{-1}}$}}
\newcommand{\gcc}{\mbox{${\rm g~cm^{-3}}$}}
\def\ph{\textcolor{black}}
\def\mg{\textcolor{magenta}}
\def\dy{\textcolor{black}}



\title[Bulk viscosity in a neutron star mantle]{Bulk viscosity in a neutron star mantle}

\author[D. G. Yakovlev et al.]{
    {  D. G. Yakovlev$^1$\thanks{E-mail: yak@astro.ioffe.ru}},
     {M. E. Gusakov$^{1}$},
    P. Haensel$^{2}$\\
    $^{1}$ Ioffe Institute, Politekhnicheskaya 26, St~Petersburg 194021, Russia\\
    $^{2}$ Copernicus Astronomical Center, Bartycka 18, 00-716 Warsaw, Poland\\
    }

\maketitle \label{firstpage}

\begin{abstract} We study the bulk viscosity in two 
(anti-spaghetti and Swiss cheese) phases of
nonspherical nuclei  in the
mantle of an oscillating neutron star near the boundary with the stellar core. The bulk
viscosity is produced by non-equilibrium 
Urca neutrino emission processes. In the mantle, the
direct Urca process may be open (Gusakov et al., 2004) if neutrons and
protons move in a periodic potential created by a lattice of non-spherical
nuclei (which allows the nucleons to have large
quasi-mo\-men\-ta and satisfy direct Urca momentum-conservation).
This bulk viscosity can dominate over
that due to the modified Urca process in the outer stellar core
and over the shear viscosity. The bulk viscosity depends strongly
on temperature, oscillation 
frequency and nucleon superfluidity. The 
enhanced bulk viscosity in the mantle can control propagation and 
damping of neutron star oscillations.

\end{abstract}



\section{Introduction}
\label{introduction}

It is widely thought \citep{ST1983} that
any neutron star contains a massive core of superdense matter
surrounded by a thin crust ($\sim$1 percent by mass and
$\sim 10$  {percent} by radius) of matter of subnuclear density. The
core is supposed to be liquid and contain neutrons (n), protons (p),
electrons (e), muons ($\mu$), and possibly other particles and
elementary excitations like hyperons, $\Delta$ isobars, deconfined
quarks. The crust is composed of electrons and atomic nuclei as well as
of free neutrons (at densities $\rho$ exceeding
the neutron drip density
$\rho_{\rm drip} \sim (4 - 6) \times 10^{11}$ g~cm$^{-3}$). In the
bulk of the crust the electrons are strongly degenerate and relativistic,
the nuclei are nearly spherical and neutron rich,
and the free neutrons are strongly degenerate and possibly superfluid. 
The stellar core also consists of highly
degenerate and strongly interacting fermions (like n and p) 
which can be superfluid.

The core-crust interface is placed at $\rho_{\rm cc} \approx \rho_0/2$ 
(e.g., \citealt{HPY2007}) which
corresponds to the baryon number densities $\nb= n_{\rm cc}\approx
0.5 \, n_0$, where $\rho_0\approx 2.8 \times 10^{14}$ g~cm$^{-3}$ and
$n_0=0.16$ fm$^{-3}$ are, respectively, the density and the baryon number density
of symmetric nuclear matter at saturation.
It is important that many thermodynamic, kinetic and neutrino-emission properties
of the crust and core are drastically different. Accordingly, the crust-core
interface is very  significant for modeling of many processes in neutron stars such
as magnetic field and thermal evolution, pulsar glitches, stellar
oscillations.

An additional complication is introduced by a possible existence
of a special nuclear pasta layer (called also nuclear mantle and
viewed often as the bottom layer of the crust) in the density range
from $\rho \sim 10^{14}$ g~cm$^{-3}$ to $\rho_{\rm cc}$. 
This layer appears in some theoretical models of
neutron star matter but is not realized in other models 
(e.g., \citealt{Ravenhalletal83,PR1995,HPY2007}). 
In the essence, the atomic nuclei in the nuclear pasta form exotic 
structures, with the sea of free neutrons (and even free protons)
between them. The bulk properties of the nuclear pasta
layer (e.g., the equation of state, EOS) remain almost the same as in
a fluid of neutrons with the admixture of protons and electrons
(almost as in the outer liquid core). 
Accordingly, the nuclear pasta \ph{weakly affects 
hydrostatic structure of the star but 
can much stronger affect transport and neutrino emission properties of the 
mantle (e.g. \citealt{LRP93,Gusakov2004,hb08,ns18} and references therein)}. 
Although the pasta layer occurs in a narrow density interval,
its mass is not much smaller than the total mass of the crust.

In this paper, we focus on the bulk viscosity in the nuclear pasta
using the results of \citet{HLY2000} and \citet{Gusakov2004}. The bulk viscosity
is required for modeling neutron star  oscillations or propagation of waves generated
in neutron stars during active periods of their evolution. In particular, 
it is
expected to be important at the stages at which neutron stars become powerful
sources of gravitational waves (e.g., in rapidly  rotating stars suffering
r-mode instabilities or during neutron star merging in compact binaries; \ph{see \citealt{Alford2018}
and references therein}).

\section{Nuclear pasta and direct Urca process}
\label{s:pasta}

\ph{Nuclear pasta in neutron stars is extensively studied using two major approaches.}

\ph{First, one considers ordered nuclear
clusters, which can exist in four exotic phases I--IV, by minimizing the energy density at a given baryon density. The phase I of rods (`spaghetti') appears  after the
standard phase of spherical nuclei as $\rho \to \rho_{\rm cc}$. Nuclear structures become
rod-like there; they are crystallized in a two-dimensional lattice immersed in a
liquid of free neutrons.  
One studies neutron and
proton density profiles across the rods assuming that their length
is infinite. The phase of rods is followed by the phase II
of slabs (the `lasagna' phase). The slabs are filled by nuclear matter
and form a one-dimensional lattice, with the neutron liquid between
the slabs. One models the nuclear structure across the slabs treating them as infinite. The next phase
III is a two-dimensional lattice of infinite rod-like bubbles of
neutron liquid (possibly with some amount of free protons) with nuclear structures between them; it is
called `anti-spaghetti.' The final phase IV consists of a crystal of spherical bubbles
of free neutrons (and possibly free protons) called `Swiss cheese';
nuclear structures fill the space between the bubbles.}

\ph{These structures
are studied  
using compressible liquid drop
models  
or Thomas-Fermi theories (e.g.,
\citealt{WK85,Oya1993,LRP93,WIS00,WATANABE-IIDA03,
NAKAZATO09,OKAMOTO2013,GRILL2014,Sharma2015}, and references therein).  Some calculations
reproduce all phases I--IV; others reproduce fewer exotic phases or
none of them (e.g. \citealt{DH2001,GRILL2014}). Of course, real rods or slabs 
should have 
finite sizes; they are possibly arranged in domains. However, the sizes of clusters and structure
of domains cannot be obtained from such calculations.}

\ph{Second, one simulates nuclear pastas using 
molecular dynamics (e.g., \citealt{MARUYAMA98,KIDO2000,WATANABE2002,
SCHUETRUMPF15,HOROWITZ04,HOROWITZ2005,DORSO2012,CAPLAN2015,
DISORDER15} and references therein). These simulations are sensitive to
the models of interacting nucleons, but 
they have more degrees of freedom. 
Accordingly, they predict nuclear structures of different sizes, shapes and orientations. One obtains a rich variety of
pasta models, for instance, nuclear clusters of single type (e.g.,
slabs) with different parameters for a given density of the matter; 
mixtures of cluster types (e.g., slabs and rods);
even more exotic clusters, called nuclear waffles,  gyroids, intertwined lasagna, 
long lived topological defects and so on (e.g.,
\citealt{SCHNEIDER14,NAKAZATO09,SCHUETRUMPF15,ALCAIN14,DISORDER15}). At 
temperatures $T \lesssim 10^{10}$ K of our interest, such
nuclear pastas  usually freeze into ordered or disordered systems.}

\ph{All these studies give rather diverse results -- wealth of possibilities.}
To be specific, we employ the well-known model I of \citet{Oya1993}.
It is the nuclear pasta model of the first type containing
all four basic pasta layers I--IV. The phase I appears at
density $\rho_1=0.973 \times 10^{14}$ $\gcc$ (at baryon number
density $n_1$=0.0586 fm$^{-3}$). The  phase II occurs at $\rho_2=1.24
\times 10^{14}$ $\gcc$ ($n_2=0.0749~{\rm fm}^{-3}$), and the phase
III at $\rho_3=1.37 \times 10^{14}$ $\gcc$ ($n_3=0.0827~{\rm
fm}^{-3}$). The final phase IV appears at $\rho_4=1.42 \times
10^{14}$ $\gcc$ ($n_4=0.0854~{\rm fm}^{-3}$) and disappears at the
crust-core interface ($\rho_{\rm cc}=1.43 \times 10^{14}$ $\gcc${,}
$n_{\rm cc}=0.0861~{\rm fm}^{-3}$). 

\ph{In the phases III and IV, free (\ph{unbound})
protons appear in a liquid of free neutrons. It happens not only
in the models by \citet{Oya1993} but in some other models 
(see, e.g., \citealt{Sharma2015}).  It is
favorable for opening
a powerful direct Urca process of neutrino emission  in the
neutron star mantle}.  
The idea was put forward by \citet{LRP93}
and realized by \citet{Gusakov2004}. 

It is well known \citep{1991DURCA} that the direct Urca process can produce the
strongest neutrino emission in the inner cores of neutron stars, at
$\rho \gtrsim (2-3)\, \rho_0$. The simplest direct Urca process
involves neutrons,  protons and electrons; it is a chain of two 
(direct and inverse) reactions,
\begin{equation}
{\rm n \to p + e} + \bar{\nu}_{\rm e}, \quad {\rm p + e \to n +}
\nu_{\rm e},
\label{durca}
\end{equation}
where $\nu_{\rm e}$ and $\bar{\nu}_{\rm e}$ are the electron
neutrino and antineutrino, respectively. In a strongly degenerate
matter the momenta of neutrons, protons and electrons should be
close to their Fermi momenta $\pFn, \pFp$, and $\pFe$. The process
is allowed by momentum conservation if $\pFn < \pFp + \pFe$, and
forbidden otherwise. Typical neutrino momenta $p_{\nu} \sim k_{\rm
B}T/c$ are much smaller than these Fermi momenta and can be
neglected in momentum conservation. Because of momentum conservation
restriction, the direct Urca process in the neutron star core is
allowed only at rather high densities (typically, a few times of
$\rho_0$) for those model EOSs of dense matter which
have a large symmetry energy (rather high fractions of protons and
electrons) to satisfy the triangle inequality $\pFn < \pFp + \pFe$. In the
outer core of the star, this process is forbidden, and only
weaker processes like the modified Urca one can operate (e.g.
\citealt{Yak2001}), with the direct and inverse reactions
nN$\to$peN$\bar{\nu}_{\rm e}$ and peN$\to$nN$\nu_{\rm e}$, N being a
nucleon.  

However, crystalline lattices of nuclear structures in the
phases III and VI of the neutron star mantle modulate motion of free
neutrons and protons (inducing Bloch states and relaxing momentum
conservation due to Bragg diffraction). This can open the direct
Urca process owing to Umklapp transitions of n and p.
The Umklapp transitions make the direct Urca process in the mantle
weaker than in the inner core but it can still be stronger than the
modified Urca processes in the outer core of neutron stars.
\citet{Gusakov2004} calculated the neutrino
emissivity of the direct Urca processes in the phases III and IV of
the nuclear pasta using two models for evaluating the matrix elements
of the processes. The results appeared  {to be quite similar} and were
fitted by an analytic expression. They will be used for calculating
the bulk viscosity.

\section{Bulk viscosity}
\label{s:bulkvisca}

Let us consider the bulk viscosity in the stellar mantle and
compare it with the bulk viscosity in the outer core. We will
mainly neglect the effects of neutron and proton superfluidities
but comment on these effects whenever necessary.

\subsection{Stellar core}
\label{s:bulkvisca-core}

The bulk viscosity $\zeta$ has been studied attentively for the
conditions in neutron star cores. Its nature is very different from
that for the standard shear viscosity $\eta$ determined by frequent
collisions between the particles in dense matter \citep{FI1979}.

A noticeable bulk viscosity in neutron star cores is produced by
slow  {irreversible} beta processes. For simplicity, we consider a matter of
neutrons, protons and electrons in an outer core of the star near
the crust-core interface. Let us assume that the matter is initially
in full thermodynamic equilibrium, including beta-equilibrium. Then
the imbalance of the chemical potentials $\delta \mu=\mu_{\rm
n}-\mu_{\rm p}-\mu_{\rm e}$ is zero.  At the next step 
we consider weak vibrations of the star with
 {a local oscillation frequency $\omega$}. 
They can be produced by global
oscillations of the star or by a propagation of  {some} sound waves. Frequent collisions of the particles
will support local thermodynamic equilibrium but Urca processes are
too slow to restore beta equilibrium during oscillations.
Oscillations of particle number densities and the pressure $P$
will be slightly shifted in phase, producing
dissipation that can be described
by an effective bulk viscosity $\zeta$. 

The first calculations of
$\zeta$ have been conducted by \citet{Sawyer1989,HS1992}; we will follow
the consideration by \citet{HLY2000,HLY2001}. 
The bulk viscosity is controlled by non-equilibrium (direct and/or modified)
Urca processes and depends on the  oscillation 
frequency $\omega$. For simplicity, we restrict ourselves by the so
called linear regime, where the amplitude $|\delta \mu|$
of $\delta \mu$ oscillations is smaller than $\kB T$, $\kB$ being the Boltzmann constant.
Note that neutron star oscillations in the non-linear regime 
($|\delta \mu| \gg \kB T$) were studied
e.g., by \citet{fw68,gyg05,ams12}.
In the linear regime, the 
difference of the rates $\Delta \Gamma$ of the direct and inverse
Urca reactions is linear in $\delta \mu$,
\begin{equation}
     \Delta \Gamma= - \lambda \delta \mu,
\label{e:DGamma}
\end{equation}
where $\lambda$ is a factor that can be expressed through the rate
$\Gamma_0$ of equilibrium Urca processes. For the
direct Urca process \citep{HLY2000},
\begin{equation}
      |\lambda|=\frac{17\pi^4}{60}\,\frac{\Gamma_0}{\kB T},
 \label{e:lambda}
\end{equation}
with $\Gamma_0$ given by Eq.\ (28) of \citet{HLY2000}. In this case
$\lambda \propto T^4$. The expressions for $\lambda$ mediated by the
modified Urca processes are given in \citet{HLY2001}, with $\lambda
\propto T^6$. If neutrons and/or protons are superfluid, neutrino reaction
rates are reduced due to the appearance of gaps in the energy spectra
of nucleons (e.g. \citealt{Yak2001}). Then the contribution
of any Urca process has to be multiplied by a corresponding reduction 
factor (e.g., \citealt{HLY2000,HLY2001}). The reduction
factors can strongly (exponentially) depend on $T$, 
violating
simple power-law temperature dependence of $\lambda$. 
If several Urca processes operate at once, one should
sum corresponding reaction factors~$\lambda$. Since $\lambda$ is sometimes
introduced with different signs we use $|\lambda|$ to avoid
misunderstandings.

In the npe-matter, the effective bulk viscosity can be written as
\citep{HLY2000}
\begin{equation}
    \zeta=\frac{C_{\rm e}^2 n_{\rm b}^2 |\lambda|}{B_{\rm ee}^2|\lambda|^2+\omega^2n_{\rm
    b}^2},
    \label{e:zeta}
\end{equation}
where $n_{\rm b}$ is the baryon number density at equilibrium, while
\begin{equation}
    C_{\rm e}=-\frac{1}{n_{\rm b}}\,\frac{\partial P(n_{\rm b},X_{\rm e})}{\partial X_{\rm
    e}}, \quad B_{\rm ee}=n_{\rm b}\, \frac{\partial \delta \mu(n_{\rm b},X_{\rm e})}{\partial n_{\rm b}},
    \label{e:CB}
\end{equation}
are the derivatives of thermodynamic pressure and chemical potential
imbalance with respect to $X_{\rm e}$ (the ratio of
the numbers of electrons and baryons) and $\nb$. The derivatives have to be
calculated by considering $n_{\rm b}$ and $X_{\rm e}$ as independent
variables and taken at equilibrium. To a good approximation, they are
independent of $T$. 

For expected vibration frequencies $ {\omega\sim 10^3 -10^4}$ s$^{-1}$ the
high-frequency limit of (\ref{e:zeta})  is an excellent
approximation,
\begin{equation}
    \zeta=\frac{C_{\rm e}^2 |\lambda|}{\omega^2}.
    \label{e:zeta-dyn}
\end{equation}
This bulk viscosity decreases as $\omega^{-2}$ with growing
$\omega$. In a non-superfluid matter, it increases with
the growth of $T$.

In the unrealistic low-frequency limit ($\omega \to 0$) the bulk
viscosity becomes static and huge,
\begin{equation}
    \zeta=\frac{C_{\rm e}^2 n_{\rm b}^2}{B_{\rm ee}^2|\lambda|}.
\label{e:low-freq}		
\end{equation}
Its dependence on $|\lambda|$ is inverted, inverting the temperature
dependence of $\zeta$. 

\ph{A transition from the  
high-frequency to the low-frequency regime occurs at 
$\omega=\omega_0 \sim |B_{\rm ee} \lambda|/n_{\rm b}$}. 
\ph{Some estimates of $\omega_{0}$  are given in Section \ref{s:illustrat}.
The estimates of $\lambda$ and $C_{\rm e}$ can be found in \citet{HLY2000,HLY2001}; 
$B_{\rm ee}$ is negative, with $|B_{\rm ee}| \sim $ a few GeV.}	  

\subsection{Pasta layers}
\label{s:bulkvisca-pasta}

The consideration of Section \ref{s:bulkvisca-core} implies that the
bulk viscosity in nuclear pasta is given by the same Eq.
(\ref{e:zeta}) with the corresponding value of $\lambda$. A
comparison of the results of \citet{HLY2000} and \citet{Gusakov2004}
indicates that the factor $\lambda$ in the pasta layers III and IV 
is given by the same expression (\ref{e:lambda}) 
corrected for the reduction of the direct Urca process due
to the presence of nuclear structures,
\begin{equation}
      |\lambda|=\frac{17\pi^4}{60}\,\frac{\Gamma_0}{\kB T}\,{\cal R}.
 \label{e:Gamma0}
\end{equation}
Here ${\cal R}={\cal R}(\nb)$ is the reduction factor given by Eq. (12) of
\citet{Gusakov2004} who approximated it at $n_3 \leq \nb \leq n_{\rm
cc}$ by their Eq.\ (15),
\begin{eqnarray}
  {\cal  R}(\nb)&=& {\cal R}_2+({\cal R}_1-{\cal R}_2)(1-x)^2\quad x<1,
\nonumber \\
  {\cal R}(\nb)&=& {\cal R}_2/x^7 \quad x \geq 1,
\label{e:R}
\end{eqnarray}
where ${\cal R}_1=6 \times 10^{-5}$, ${\cal R}_2=10^{-5}$, and
$x=(\nb-n_3)/(n_4-n_3)$, with $n_3=0.0827~{\rm fm}^{-3}$,
$n_4=0.0854~{\rm fm}^{-3}$ and $n_{\rm cc}=0.0861~{\rm fm}^{-3}$.
If free nucleons in the pasta are superfluid, Eq.\ (\ref{e:Gamma0}) 
should be multiplied by the additional superfluid reduction factor, which
is the same as discussed by \citet{Gusakov2004}. 
Equations (\ref{e:zeta}), (\ref{e:Gamma0}) and (\ref{e:R}) allow one
to evaluate $\zeta$ in the pasta layers.

\subsection{Illustrative examples}
\label{s:illustrat}

\begin{figure*}
\includegraphics[width=0.47\textwidth]{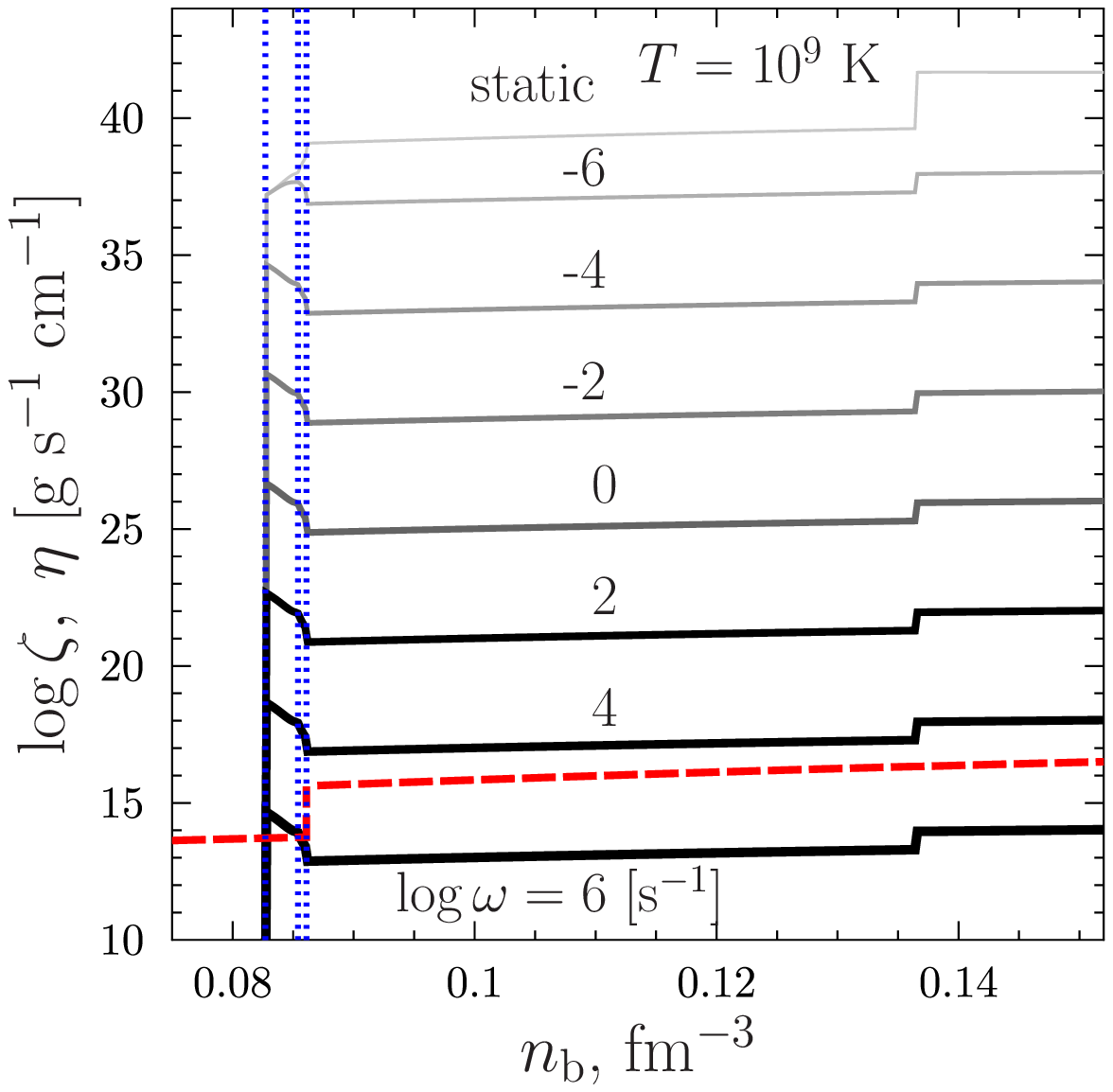}%
\hspace{5mm}
\includegraphics[width=0.47\textwidth]{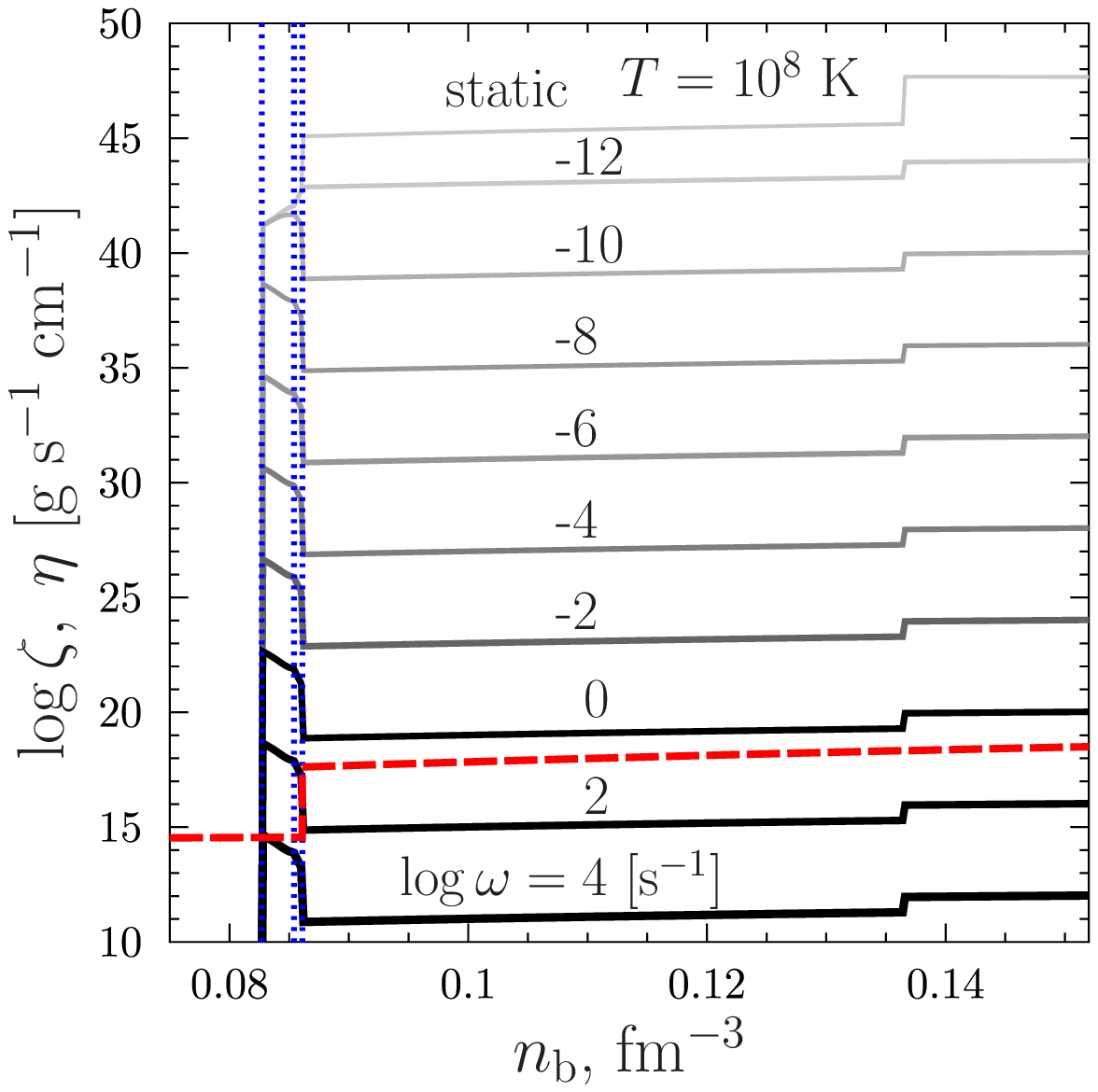}%
\caption{(Color online) Bulk viscosity $\zeta$ (solid lines)  versus
baryon number density $\nb$ in the pasta layers III and IV 
and in the outer layer of the core at $T=10^9$ K (left) and $10^8$ K (right).
Lines of different thickness and grayness refer to different vibration
frequencies $\omega$ (the values of $\log \omega$ are given near the curves); the upper
light gray lines show the static $\zeta$ (at $\omega=0$). For comparison,
the long-dashed lines demonstrate the shear viscosity $\eta$. The left dotted
vertical line in each panel shows the outer boundary of the anti-spaghetti pasta
layer III; the next vertical line positions the onset of the Swiss-cheese
layer IV, and the right vertical line marks the outer boundary
of the stellar core. 
The effects of superfluidity are neglected.
See text for details.
}
\end{figure*}

Fig.\ 1 illustrates the behavior of the bulk viscosity $\zeta$ (solid
lines) as a function of $\nb$ 
in the pasta layers III and IV and in the outer
layer of the core (neglecting the effects of superfluidity). 
The left-hand panel is for the temperature
$T=10^9$ K, while  {the right-hand panel} is for $T=10^8$ K.
The pasta phase III occurs between the left and middle dotted vertical
lines while the phase IV is realized between the middle and right dotted vertical lines.
Higher $\nb$ correspond to the stellar core. Different solid lines are plotted
for different oscillation frequencies $\omega$; the upper solid line 
shows the static $\zeta$ at $\omega \to 0$.

In the core, we use a simple model  
EOS of npe$\mu$-matter suggested by \citet{HHJ1999}. 
Specifically, we use the same version of this EOS
as employed by \citet{KAM2014} who called it HHJ(0.1,0.7). Its
advantage is that one can easily calculate
the thermodynamic derivatives (\ref{e:CB}).
For this EOS, the maximum gravitational mass of stable neutron stars
is  $M_{\rm max}=2.16\,{\rm M}\odot$. The central density of 
the maximum-mass star is
$\rho=2.45 \times 10^{15}$ $\gcc$ ($\nb=1.06$ fm$^{-3}$), 
and the circumferential radius
$R=10.84$ km. The direct Urca processes operate in stars with
$M>1.77\,{\rm M}\odot$ (with central densities $>1.05 \times 10^{15}$
$\gcc$ and $\nb>0.558$ fm$^{-3}$). The muons appear in the core at 
$\rho=2.32 \times 10^{14}$ $\gcc$ ($\nb=0.137$ fm$^{-3}$), just
near the outer core boundary. Their appearance induces the 
modified Urca process with muons and slightly increases $\zeta$. 
The nuclear pasta model in Fig.\ 1 is the same
model by  \citet{Oya1993} as described in Section \ref{s:pasta}.
The bulk viscosity we discuss operates in the pases III and IV. 

Thermodynamic derivatives (\ref{e:CB}) in the core are
calculated in the same manner as in \citet{HLY2000,HLY2001}. The bulk
viscosity in the outer npe-layer of the core is computed  using
Eq.\ (\ref{e:zeta}). For deeper layers, containing
muons, we have used Eq. (9) of \citet{HLY2000}. 

The thermodynamic derivatives (\ref{e:CB}) in the pasta layers are,
in principle, affected by nuclear structures. We have tried to 
include this effect by using the so called bulk approximation
described, e.g. in \citet{HPY2007}, Section 3.4.1. However,
because the pasta layers III and IV are very close to the core,
the effect is not large, and, for simplicity, we determine the required derivatives
in the same way as in the outer npe-core. \ph{Note that the factor ${\cal R}$ in (\ref{e:R}) is obtained
\citep{Gusakov2004}  for the model of 
\citet{Oya1993}. This means that the curves in Fig.\ 1 are plotted for a mixture of EOSs of dense matter. It is a typical situation in neutron star physics, that different layers of the star are described by different EOSs. Regretfully, we have no data to calculate
$\zeta$ in the core using the results by \citet{Oya1993}. Equally, we cannot accurately calculate $\zeta$ in the mantle using the HHJ EOS because we have no data on nuclear pasta for that EOS. Nevertheless, as always happens in such cases, we can expect that Fig. 1 is qualitatively correct.  
}

For comparison, by long-dashed lines in Fig.\ 1 we present also the shear viscosities
$\eta$ in the innermost layer of the crust and in the outermost core layer.
In the core, we plot $\eta$ \ph{using the same HHJ EOS} and results of \citet{PS2008}. 
In the crust, we employ the results by \citet{Chugunov2005,PS2008a}. 
In the latter case we use the smooth composition model of the
inner crust (\citealt{HPY2007}, Appendix B.2.1) neglecting the presence
of nuclear pasta. 
Note that the shear viscosity in the
pasta layers was studied 
by \citet{hb08, ns18}, who showed that
it does not differ drastically from $\eta$ 
calculated for spherical nuclei.
In contrast to $\zeta$, the shear viscosity is independent of $\omega$.

As seen from Fig.\ 1, the bulk viscosity is very sensitive to the
temperature and  oscillation frequency $\omega$. The expected  
oscillation 
frequencies are $ {\omega \sim 10^3-10^4}$ s$^{-1}$, but we consider
a much wider range of $\omega$ for the completeness of our study. 
Typically, the high-frequency
regime  {(\ref{e:zeta-dyn})} is seen to be an excellent approximation, so that
$\zeta \propto T^4/\omega^2$ in the pasta and $\zeta \propto T^6/\omega^2$
in the outer core (Sections \ref{s:bulkvisca-core} and \ref{s:bulkvisca-pasta}). 
Accordingly, $\zeta$ decreases with decreasing $T$ 
(e.g. during neutron star cooling)
and increasing $\omega$. The bulk viscosity is especially effective
in warm neutron stars, where Urca processes are stronger. In any case 
the bulk viscosity in the outer core (powered by the modified Urca process) 
is smaller than in the pasta, where it is powered by the direct Urca process
partially suppressed by the crystalline structure effects \citep{Gusakov2004}.
In the inner cores of neutron stars the direct Urca processes are stronger
and $\zeta$ can be much larger. This is not shown in Figure 1 but is well known
(e.g., \citealt{HLY2000,HLY2001}).

According to Fig.\ 1, at any temperature and density there exists some 
minimum oscillation frequency $\omega_{*}$ above which the bulk viscosity
$\zeta$ becomes smaller than the shear viscosity $\eta$. For instance,
at $T=10^9$ K in the pasta $\omega_{*}\sim 10^6$ s$^{-1}$, while
in the outer core $\omega_{*} \sim 3 \times 10^4$ s$^{-1}$.    
At $T=10^8$ K we have $\omega_{*}\sim 10^2$ s$^{-1}$ in the pasta
and $\omega_{*} \sim 3$ s$^{-1}$ in the outer core. 

\ph{Let us recall that we use one EOS, HHJ(0.1,0.7), in the
neutron star core. A comparison with the results by \citet{HLY2000,HLY2001},
who used two other, sufficiently different EOSs, 
reveals that at the same $\omega$ and
$T$ the values of $\zeta$ for different EOSs are close (differ by 
a factor of few). The largest difference occurs at densities of muon appearance
and at threshold densities for opening Urca processes. It seems that by varying
the EOS in the non-superfluid core one cannot noticeably change the values 
of $\zeta$. The same is so for the values of $\eta$ and $\omega_*$.} 

\ph{With decreasing $\omega$, the transition to the static $\zeta$, 
Eq.\ (\ref{e:low-freq}), takes takes place at unrealistically low
frequencies, $\omega_0 \sim 10^{-6}$ s$^{-1}$ for $T=10^9$ K 
and $\omega_0 \sim 10^{-12}$ s$^{-1}$
for $T=10^8$ K}. As already mentioned above, the static $\zeta$ has the inverted temperature 
dependence with respect to the high-frequency $\zeta$. The static $\zeta$
in the pasta becomes lower than in the outer core, and the static $\zeta$
increases with decreasing $T$.

If nucleons are superfluid, one should calculate $\zeta$ by introducing
superfluid reduction factors in the quantities
$\lambda$ as discussed above. For instance, in the
high-frequency limit superfluidity always suppresses $\zeta$ in exactly the same
manner as discussed by \citet{HLY2000,HLY2001} for the case of stellar core.
The suppression of $\zeta$ in the pasta is the same as the reduction of
corresponding neutrino emissivity of the direct Urca 
process analyzed by \citet{Gusakov2004}.

\ph{Therefore, in contrast to the effects of EOS, which affect $\zeta$
rather slightly, the effects of superfluidutity of neutrons and protons are really
dramatic. The most important parameters there are critical temperatures of
neutrons and protons as functions of density (e.g. \citealt{2014SF} and references therein). 
They are
extremely model-dependent, very sensitive to in-medium effects in dense matter. The region of
the crust-core interface is just the place where theoretical models for superfluidity
are very divergent. If superfluidity in the mantle is strong it can completely
suppress high-frequency $\zeta$ there 
(as follows from the results by \citealt{Gusakov2004}). If superfluidity is stronger in the core,
it will suppress the bulk viscosity in the core \citep{HLY2001}. Another important effect of superfluidity
on the bulk viscosity is described in the next section.}

\section{Discussion}
\label{s:discussion}

The effective bulk viscosity considered above is well known in 
hydrodynamics of liquids in which
slow chemical reactions are possible (e.g., \citealt{LL1987}).
It produces viscous heat to be included 
in the equation of heat transport and effective viscous force in the
Navier-Stokes equation. In vibrating neutron stars, such a bulk viscosity 
appears because vibrations violate beta equilibrium
and induce slow (weak) non-equilibrium 
processes which drive the system to the full equilibrium. Such processes are
not necessarily associated with the emission of neutrinos but can be governed, for instance,
by weak reactions between hyperons (e.g., \citealt{GK2008}). 

Let us stress that there is also a standard bulk viscosity mediated
by frequent interparticle collisions (e.g. \citealt{PhysKin1981}). It is quite similar to traditional
transport coefficients like the shear viscosity or thermal conductivity.
In particular, it is independent of  oscillation 
frequency and can be used for
any (not only pulsating) motions. However, under
the neutron star conditions this 
`standard' bulk viscosity seems mainly negligible. The frequency dependence 
in our case occurs because non-equilibrium beta processes can be much slower 
than  oscillations. 

\dy{The calculated bulk viscosity is valid under restricted
conditions. 
First of all, the variations of the chemical potential
imbalance, by construction (e.g. \citealt{HLY2000}), have to be small, $|\delta \mu| \ll \kB T$. 
Secondly, such a bulk viscosity is obtained by averaging over 
one pulsation cycle (see, again, \citealt{HLY2000}). 
Note that in the static limit ($\omega \to 0$) 
the latter approximation can be 
violated, and the problem should be considered more carefully. However,
under the neutron star conditions, this limit is unrealistic. Accordingly, 
we use the available expressions at $\omega \to 0$ just to show 
the variation 																																											
amplitude of bulk viscosity values. 
Thirdly, for the presented expressions to be valid,
the matter should oscillate as a whole, whereas
sometimes several matter components oscillate differently. This is especially
so if neutrons and/or protons are superfluid; such superfluid components
of the matter may move separately which naturally leads  
to													several coefficients of bulk  viscosity (e.g., \citealt{gusakov07,GK2008}). 
Finally, matter motions
are not necessarily reduced to pure oscillations at a fixed frequency $\omega$.
The motions can be complicated and the velocity can be an arbitrary function
of  time $t$, irreducible to oscillations with fixed $\omega$.} 

In all these cases, the relaxation of stellar perturbations due to 
non-equilibrium beta processes can be taken into account by
introducing a system of hydrodynamic equations for different components
of the matter  (see, e.g., \citealt{gyg05,kg09} for an example of such calculations). 
Various quantities in these
equations (for instance, chemical potentials) can be out of equilibrium and
have to be treated as functions of a local baryon number density $\nb$
and local instantaneous particle fractions. A solution of such a system 
automatically includes all the effects described by the
phenomenon of bulk viscosity. This approach can also be
valid in the non-linear regime in which  {$|\delta \mu| \gg \kB T$} and
beta-equilibration is greatly accelerated by  strong violations from
beta-equilibrium.      

Let us recall that the bulk and shear viscosities damp different types of
motions $\bm{v}(\bm{r},t)$. The shear viscosity affects shear motions 
while the bulk viscosity influences compression-rarefaction motions with
${\rm div}\bm{v}\neq 0$. For instance, standard sound waves are damped
by both viscosities (e.g., \citealt{LL1987}). In a matter with a not
too strong magnetic fields $\bm{B}$, where $B^2/8\pi \ll P$, the fast
magnetic sound waves are  {also damped as the ordinary sound}, but
the  {Alfv\'enic} waves and slow magnetic sound waves have shear-like structure and
they are mostly damped by the shear viscosity. The damping of proper oscillations
of the star is determined by configurations of fluid motion in these
oscillations.

\section{Conclusions}
\label{s:conclude}

We have calculated the bulk viscosity in the two  
(anti-spaghetti and Swiss cheese) phases of
nuclear pasta in the mantle of vibrating neutron stars. The pasta
phases are described by model 1 of \citet{Oya1993}. The
bulk viscosity is mediated by direct Urca process which is
opened \citep{Gusakov2004} because free neutrons and protons are moving  
there in a periodic potential
created by a lattice of nuclear structures. 

We have compared
the calculated bulk viscosity in the pasta with the bulk viscosity
in the outer neutron star core and with appropriate shear viscosities.
The bulk viscosity appears to be a strong function of temperature, density
and vibration frequency $\omega$. For typical vibration frequencies
$\omega \sim 10^3-10^4$ s$^{-1}$ the bulk viscosity in the pasta exceeds the
bulk viscosity in the outer core as well as the shear viscosity in the 
core and in the crust at high enough 
temperatures  {$T \gtrsim (1.5-3) \times 10^8$} K. In addition, the bulk
viscosity can be drastically affected (mostly suppressed) by superfluidity (Section \ref{s:illustrat}).
The bulk viscosity can influence the propagation of waves through the pasta layer and damping of 
various neutron star  oscillations, as well as of  various hydrodynamical motions
in the vicinity of the pasta layer. \dy{It can also affect damping
of perturbations in a neutron star merging with a compact
companion during the emission of gravitational waves
(e.g., \citealt{Alford2018,AH2018}).}

\dy{It is important to stress that the employed pasta model \citep{Oya1993}
is just one example in a sea of many diverting
models
(Section \ref{s:pasta}). Particularly,
that model essentially requires pure crystalline lattice
of exotic nuclear structures. Even a slight lattice disorder can affect the bulk viscosity via the factor ${\cal R}$ in Eq.\ (\ref{e:R})
(through summations over lattice
vectors in Eqs. (8) and (12) of \citealt{Gusakov2004}; note also an approximate character of these summations which would be desirable to elaborate in the future). Even for pure lattice, the
bulk viscosity can be drastically changed by superfluidity. The bulk viscosities for other pasta models have not been studied in the literature.
Different pasta models should provide a wealth of speculations regarding transport and neutrino
emission properties of the pasta layers (a huge job for future
studies).} 

For the bulk viscosity (and the direct Urca process) in the mantle in our model to be strong, one needs the presence of free neutrons and protons. Note that some models
of neutron star matter predict the appearance of free protons in a neutron star
crust (near the core) even in the absence of nuclear pasta
(e.g., \citealt{BSk2013}). This may trigger additional neutrino emission 
and associated bulk viscosity at the bottom of the ordinary crust. 

\dy{As already mentioned above, molecular dynamics simulations of electron thermal and electrical conductivities and shear viscosity 
in the pasta layers (e.g. \citealt{hb08,ns18})
indicate that kinetic properties of nuclear pasta
are not essentially different from those in 
the inner crust of neutron stars with spherical nuclei. This would mean that the pasta layer is {\it not} a special layer possessing extraordinary
kinetic properties.} 

\dy{We note that the above conclusion is based on molecular dynamics models of pasta layers (Section \ref{s:pasta}), whereas we use the pasta model of other type whose kinetic properties have not been thoroughly investigated; for our pasta model, the conclusion might be different. In any case, the situation with the bulk viscosity can be different
because the bulk viscosity 
is too different from other transport coefficients. The bulk viscosity seems to be much more sensitive to the properties of  pasta layers and it can play a special role there. It is strongly
affected by superfluidity. In addition, the conclusion on the standard kinetic properties of the pasta layers, obtained from molecular dynamics, has been made neglecting the effects of strong magnetic fields, while the presence of such fields 
can drastically affect all the kinetic properties of the pasta layers that is almost unexplored \citep{PASTA}. It might be so that the magnetic fields make kinetic properties of nuclear pasta extraordinary. In particular, the electron shear viscosity in a magnetic field becomes much more
complicated (e.g., \citealt{OfYak2015}) than the thermal or electrical conductivities.}   

\dy{Finally, in less deep crustal layers the problem of beta equilibrium
is also important. In the essence, beta equilibrium is mediated by
beta processes on highly unusual, neutron rich nuclei (direct and inverse
reactions of Urca processes on nuclei). Such processes can be extremely slow and keep the
crust out of beta equilibrium for a very long time except for those places 
in the crust where Urca pairs of atomic nuclei coexist (e.g. \citealt{Imshennik1968,BSS1972,Deibeletal2016},
and references therein) which 
could strongly accelerate beta-equilibration.} 

\dy{In summary, there are many properties of neutron star crust and mantle,
which need to be explored that is definitely out of the scope of the present paper}. 

\section*{Acknowledgments}
 {The work of PH was supported by the 
Polish National Science Centre (NCN)  (grant no. 2013/11/B/ST9/04528),
the work of MG by the Foundation for the Advancement of Theoretical Physics BASIS 
[Grant No. 17-12-204-1 (MEG)], and the work
of DY was partly supported by the Russian
Foundation for Basic Research (grant 16-29-13009-ofi-m). }

\bibliographystyle{mnras}


\end{document}